\documentclass[aps,prb,twocolumn,amsmath,amssymb,showpacs,superscriptaddress,notitlepage,longbibliography]{revtex4-2}
\usepackage[colorlinks=true,linkcolor=blue,anchorcolor=red,citecolor=blue, urlcolor=blue]{hyperref}
\usepackage{bm}
\usepackage{graphicx}
\usepackage{color}
\usepackage{makecell}
\usepackage{multirow}
\newcommand{\Tr}{\operatorname{Tr}}
\newcommand{\sgn}{\operatorname{sgn}}
\newcommand{\tb}{\textbf}
\newcommand{\tr}{\textrm}
\newcommand{\mc}{\mathcal}
\begin{document}
\title{Excess photon-assisted noise of Majorana and Andreev bound states}

\author{Qiang Lin}
\affiliation{Beijing Academy of Quantum Information Sciences, Beijing 100193, China}
\affiliation{Beijing National Laboratory for Condensed Matter Physics, Institute of Physics, Chinese Academy of Sciences, Beijing 100190, China}
\affiliation{University of Chinese Academy of Sciences, Beijing 100049, China}

\author{Ying-Xin Liang}
\affiliation{Beijing Academy of Quantum Information Sciences, Beijing 100193, China}

\author{Ke He}
\affiliation{State Key Laboratory of Low Dimensional Quantum Physics, Department of Physics, Tsinghua University, Beijing 100084, China}
\affiliation{Beijing Academy of Quantum Information Sciences, Beijing 100193, China}
\affiliation{Frontier Science Center for Quantum Information, Beijing 100084, China}
\affiliation{Hefei National Laboratory, Hefei 230088, China}

\author{Zhan Cao}
\email{caozhan@baqis.ac.cn}
\affiliation{Beijing Academy of Quantum Information Sciences, Beijing 100193, China}

\begin{abstract}
Photon-assisted tunneling arises under an ac bias, with the drive frequency setting the photon energy. The excess photon-assisted noise is defined as the difference between the shot noise under a combined dc and ac bias and that under a dc bias alone. We investigate this quantity in tunneling into Majorana or Andreev bound states, which are of great interest in the search for topological superconductors. Under a harmonic bias $V(t)=V_\mathrm{dc}[1-\cos(\Omega t)]$, the excess photon-assisted noise exhibits distinct behaviors: for Majorana or quasi-Majorana bound states, it undergoes multiple sign reversals as $V_\mathrm{dc}$ increases and vanishes at nonzero integer values of $eV_\mathrm{dc}/\Omega$ (with $e$ the elementary charge), whereas for zero-energy Andreev bound states—particularly those producing nearly quantized zero-bias conductance peaks—it remains strictly negative over the entire $V_\mathrm{dc}$ range.
\end{abstract}

\maketitle

\section{Introduction}\label{intro}
Majorana bound states (MBSs)~\cite{read2000paired,kitaev2001unpaired}, predicted to appear at the boundaries or vortex cores of topological superconductors~\cite{alicea2012new,sato2016majorana}, are promising for realizing topological quantum computation~\cite{kitaev2003fault,nayak2008non,sarma2015majorana}. An isolated MBS resides at zero energy as an equal-weight electron–hole superposition, enabling resonant Andreev reflection~\cite{law2009majorana} and thereby producing a quantized zero-bias conductance peak (ZBCP)~\cite{law2009majorana,flensberg2010tunneling,wimmer2011quantum}. Owing to their topological protection, MBS-induced ZBCPs are expected to remain robust against variations in experimental parameters, manifesting as stable quantized conductance plateaus.

Among the various candidates for topological superconductor~\cite{mandal2023topological}, iron-based superconductors~\cite{chen2019quantized,zhu2020nearly} and semiconductor–superconductor hybrid nanowires~\cite{wang2022plateau} have exhibited nearly quantized conductance plateaus. 
Numerical simulations, however, indicate that such plateaus may also arise from zero-energy Andreev bound states (ABSs)~\cite{sarma2021disorder,pan2021quantized} or quasi-MBSs (QMBSs)~\cite{moore2018quantized,vuik2019reproducing}; the latter reproduce local features of MBSs but lack intrinsic topological protection~\cite{vuik2019reproducing}. These topologically trivial states typically originate from disorder~\cite{liu2012zero,ahn2021estimating,sarma2023spectral,pan2020physical} or spatially varying physical parameters~\cite{kells2012near,fleckenstein2018decaying,liu2017Andreev,moore2018two,cao2019decays,stanescu2019robust,pan2020physical} in realistic hybrid nanowires. To achieve an unambiguous identification of MBSs in hybrid nanowires, the ``topological gap protocol'' has been proposed~\cite{pikulin2021protocol}, wherein electrodes are attached to opposite ends of a grounded nanowire to measure both local and nonlocal conductances. This approach tests whether ZBCPs appear simultaneously in the local conductances following a bulk gap closing and reopening observed in the nonlocal conductance—signatures expected at a topological superconducting transition~\cite{lutchyn2010majorana,oreg2010helical}. While a recent experiment on InAs–Al hybrid nanowires~\cite{aghaee2023inas} reported consistency with this protocol, the data quality was deemed insufficient for a definitive identification of MBSs~\cite{kouwenhoven2025perspective}. Moreover, the robustness and reliability of this protocol remain controversial~\cite{hess2023tivial,sarma2024comment,hess2024reply,legg2025comment}.

In superconducting vortex systems expected to host MBSs, such as topological insulator–superconductor heterostructures~\cite{fu2008superconducting} and iron-based superconductors~\cite{wang2015topological}, nonlocal transport measurements remain unrealized despite several proposed protocols~\cite{bolech2007observing,sbierski2022identifying}, likely due to experimental challenges. Instead, various STM–based local probes have been employed in these systems. For example, spin-polarized STM tips~\cite{sun2016majorana,wang2021spin} probe MBS-induced spin-dependent tunneling~\cite{he2014selective}, while superconducting tips provide tunneling spectroscopy with enhanced energy resolution compared to metallic tips~\cite{ge2023single}.

Superconducting tips have also been used for shot noise measurements in iron-based superconductors~\cite{ge2023single}. The extracted Fano factor approaches unity in the large-bias, low-temperature limit, consistent with theoretical predictions for MBSs~\cite{perrin2022identifying}. Subsequent theoretical work~\cite{cao2023differential} suggested, however, that zero-energy ABSs can exhibit the same feature. Differential shot noise has also been proposed as an identifier to distinguish zero-energy ABSs from MBSs and QMBSs, as they produce distinct noise signatures near zero bias~\cite{cao2023differential,wong2022shot}. Yet, low-bias shot noise, which scales with the bias, can be obscured when significant background noise is present~\cite{liang2022low}, thereby limiting the practical applicability of this protocol.

In this work, we investigate photon-assisted shot noise in tunneling into MBSs, QMBSs, or zero-energy ABSs under a time-dependent bias. Applying an ac bias at frequency $\Omega$ allows charge carriers to absorb or emit photons of energy $\hbar\Omega$ (hereafter we set $\hbar=1$), a process known as photon-assisted transport~\cite{platero2004photon}. Photon-assisted shot noise has previously been observed in coherent normal systems~\cite{schoelkopf1998observation,reydellet2003quantum,gasse2013observation,dubois2013minimal,gabelli2013shaping} as well as in diffusive normal-metal–superconductor junctions~\cite{kozhevnikov2000observation}, where it typically displays features at a sequence of dc bias values separated by $\Omega$ or $\Omega/2$. It is therefore natural to expect that zero-energy bound states should likewise generate distinct photon-assisted shot noise signatures at large bias, where the influence of background and/or thermal noise is substantially reduced. In practice, we specifically focus on the \emph{excess photon-assisted noise}, defined as
\begin{equation}
S^\tr{exc}(V_\tr{dc}) = S^\tr{dc+ac}(V_\tr{dc}) - S^\tr{dc}(V_\tr{dc}), \label{excessnoise}
\end{equation}
where $S^\tr{dc+ac}$ and $S^\tr{dc}$ denote the shot noise under a combined dc and ac bias and under a dc bias alone, respectively, with $V_\tr{dc}$ the dc bias voltage. Excess photon-assisted noise has proven to be a powerful probe in mesoscopic physics. In normal tunnel junctions~\cite{dubois2013minimal,gabelli2013shaping}, it has been employed to detect single-electron excitations above the Fermi sea without accompanying electron–hole pairs. For periodic Lorentzian voltage pulses of frequency $\Omega$, such excitations yield $S^\tr{exc}=0$ at zero temperature when $eV_\tr{dc}/\Omega$ equals nonzero integers~\cite{keeling2006minimal,dubois2013integer}. In normal-metal–superconductor junctions dominated by Andreev tunneling, $S^\tr{exc}=0$ is predicted when $eV_\mathrm{dc}/\Omega$ equals half-integers~\cite{belzig2016elementary,bertin2022microscopic}, reflecting the effective charge $2e$ transferred in Andreev processes. 

Here, we focus on photon-assisted Andreev tunneling mediated by the aforementioned zero-energy states in the parameter regime $k_B T \lesssim \Gamma \ll \Omega$, where $k_B T$ denotes the thermal energy and $\Gamma$ the tunneling strength. Under a harmonic bias $V(t)=V_\mathrm{dc}[1-\cos(\Omega t)]$, we find that the excess noise $S^\mathrm{exc}$ associated with MBSs or QMBSs exhibits multiple sign reversals with increasing $V_\tr{dc}$ and vanishes at nonzero integer values of $eV_\mathrm{dc}/\Omega$. By contrast, for zero-energy ABSs, particularly those yielding nearly quantized ZBCPs, $S^\mathrm{exc}$ remains strictly negative over the entire bias range. These results are supported by both effective-model analyses and numerical simulations of semiconductor–superconductor hybrid nanowires.

The remainder of this paper is organized as follows. Section~\ref{secII} introduces the effective model Hamiltonian, scattering matrix, transport formulas, and analytical results. Section~\ref{secIII} presents numerical results of the effective model together with simulations of hybrid nanowires. A summary and discussion are provided in Sec.~\ref{secIV}, and technical details are included in the Appendices.

\begin{figure}[t!]
\centering
\includegraphics[width=0.9\columnwidth]{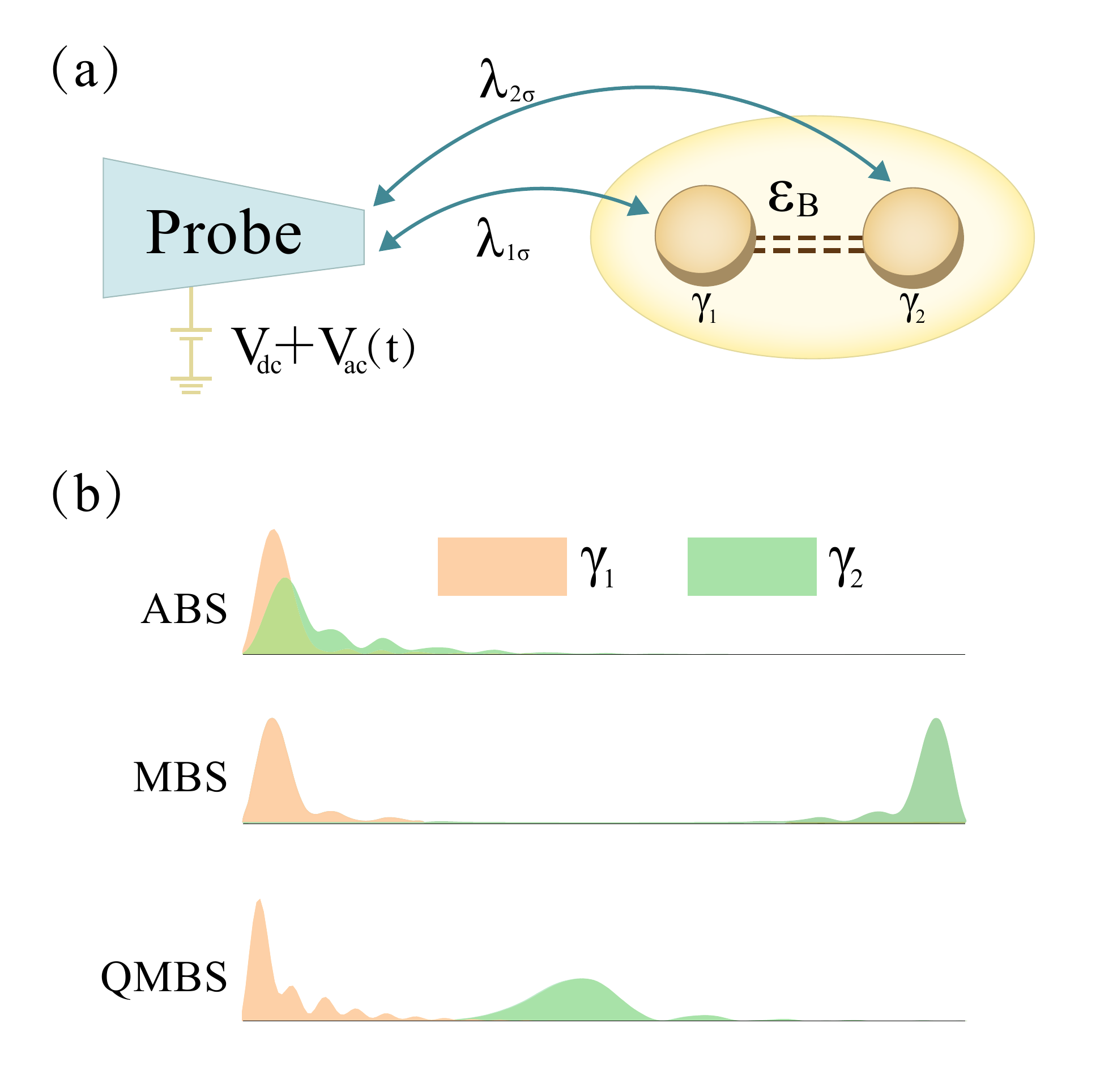}
\caption{(a) Schematic of a metallic probe tunnel-coupled to an ABS, with a bias containing both dc and ac components applied to the probe. The ac bias component enables photon-assisted tunnelings, with the drive frequency setting the photon energy. The ABS with energy $\varepsilon_B$ is represented as a pair of hybridized Majoranas, $\gamma_1$ and $\gamma_2$. $\lambda_{1\sigma}$ and $\lambda_{2\sigma}$ denote the spin-dependent tunneling amplitudes between the probe and $\gamma_{1,2}$. (b) Characteristic spatial profiles of the wave functions of ABSs, MBSs, and QMBSs in semiconductor–superconductor hybrid nanowire systems in the Majorana basis. These indicate that when $\lambda_{2\sigma}\ne 0$, the probe detects an ABS, whereas when $\lambda_{2\sigma}= 0$, it detects an MBS or QMBS. Whether $\gamma_1$ represents an MBS or a QMBS depends on the position of its partner $\gamma_2$, which, however, is inaccessible in the single-probe setup shown in (a).}\label{Fig1}
\end{figure}

\section{Model and formalism}\label{secII}
\subsection{Effective Hamiltonian}\label{secIIA}
We consider a tunneling model described by the effective Hamiltonian $H=H_P+H_B+H_T$ with 
\begin{eqnarray}
H_P&=&\sum_{k\sigma}[\varepsilon_{k}+eV(t)]c_{k\sigma}^{\dag}c_{k\sigma},\label{H_P}\\
H_{B}&=& i\varepsilon_B \gamma_1 \gamma_2, \label{H1_ABS}\\
H_{T}&=& \sum_{mk\sigma}(\lambda_{m\sigma} c_{k\sigma}^\dagger-\lambda^\ast_{m\sigma} c_{k\sigma})\gamma_{m}. \label{H1_tunneling}
\end{eqnarray}
Here, $H_P$ describes a metallic probe, such as an electrode or STM tip with dispersion $\varepsilon_k$. The fermionic operator $c_{k\sigma}$ ($c_{k\sigma}^\dagger$) annihilates (creates) an electron with momentum $k$ and spin $\sigma=\{\uparrow,\downarrow\}$. A time-dependent bias $V(t)$ is applied to the probe~\cite{jauho1994time,maciejko2006time}. The term $H_B$ describes an ABS at energy $\varepsilon_B$ deep inside the superconducting gap. It is expressed in terms of two hybridized Majorana operators, $\gamma_1$ and $\gamma_2$, which are self-conjugate ($\gamma_m^\dagger=\gamma_m$) and obey the anticommutation relation $\{\gamma_m,\gamma_n\}=\delta_{mn}$. Equivalently, the ABS Hamiltonian can be written as $\varepsilon_B c_B^\dagger c_B$, where the conventional complex fermion operator $c_B$ is related to $\gamma_1$ and $\gamma_2$ via $c_B=(\gamma_1+i\gamma_2)/\sqrt{2}$. 
$H_T$ describes the tunnel couplings between the probe and the two Majoranas. The tunneling amplitudes can be parameterized as~\cite{cao2023differential,prada2017measuring} 
\begin{equation}\label{lambda_spin}
\begin{aligned}
&&\lambda_{1\uparrow} = \lambda_{1}\sin(\theta_1/2), \qquad \lambda_{1\downarrow} = -\lambda_{1}\cos(\theta_1/2), \\ 
&&\lambda_{2\uparrow} = -i\lambda_{2}\sin(\theta_2/2), \qquad \lambda_{2\downarrow} = -i\lambda_{2}\cos(\theta_2/2),
\end{aligned}
\end{equation}
where $\theta_1$ and $\theta_2$ accounts for the spin degree of freedom of the ABS. A schematic representation of this effective tunneling model is shown in Fig.~\ref{Fig1}(a). While this model neglects Bogoliubov quasiparticles outside the superconducting gap, it captures subgap transport faithfully and allows for transparent analytical insights.

Since the probe is unpolarized, the spin quantization axis can be rotated such that $\gamma_1$ couples exclusively to the spin-down channel. We therefore set $\theta_1=0$ and retain $\theta_2$ as a free parameter. This choice is consistent with the finding that only the average angle $\theta=(\theta_1+\theta_2)/2$ enters the expressions of current and shot noise under a dc bias~\cite{cao2023differential}. For later convenience, we define the tunneling strengths $\Gamma_{1,2}=2\pi\rho\lambda_{1,2}^2$, where $\rho$ is the probe density of states at the Fermi level. Without loss of generality, we take $\Gamma_2\leq \Gamma_1$ and introduce $\Gamma=\Gamma_1+\Gamma_2$ together with a ratio 
\begin{equation}
r=\Gamma_2/\Gamma.
\end{equation}
As elucidated below, zero-energy ABSs are characterized by $r>0$, whereas both MBSs and QMBSs are characterized by $r=0$.

To connect the effective model with a representative platform, we consider an electrode tunnel-coupled to one end of a semiconductor-superconductor hybrid nanowire, which can host ABSs, MBSs, or QMBSs. All three may occur at zero energy, but they are distinguished by the properties of their wave functions in the Majorana basis. As illustrated in Fig.~\ref{Fig1}(b), the two Majorana components of an ABS strongly overlap, whereas in an MBS or QMBS they are spatially separated. QMBSs mimic MBSs near one end of the wire, which motivates their designation; however, unlike MBSs, their wave functions expressed in the electron-hole basis typically exhibit unequal electron and hole weights in the tail extending toward the opposite end~\cite{cao2022probing}. 
When the probe couples to an ABS, both Majorana components $\gamma_{1}$ and $\gamma_{2}$ couple to the probe with finite strengths $\Gamma_{1}$ and $\Gamma_{2}$, yielding $r>0$. By contrast, when the probe couples to an MBS or QMBS $\gamma_{1}$, its partner $\gamma_{2}$ is far from the probe and has negligible wave-function weight at the probed end of the wire, leading to $\Gamma_{2}=0$ and thus $r=0$. Distinguishing MBSs from QMBSs requires an additional probe at the opposite end of the wire to determine whether $\gamma_{2}$ resides there~\cite{moore2018two}. Such a two-probe configuration lies beyond the scope of the present work. The parameter $\theta$ accounts for the spinor structure of the wave functions, which is not depicted in Fig.~\ref{Fig1}(b). Since the local probe under consideration cannot distinguish between MBSs and QMBSs, we use the notation ``MBSs/QMBSs'' to denote either case hereafter. 

\subsection{Scattering matrix}\label{secIIB}
Our main findings of excess photon-assisted noise are associated with a distinctive property of the scattering matrix associated with MBSs/QMBSs and zero-energy ABSs, as elaborated below. The scattering matrix takes the general form  
\begin{equation}
\tb{s}(E)=
\begin{bmatrix}
\tb{s}^{ee}(E) & \tb{s}^{eh}(E) \\
\tb{s}^{he}(E) & \tb{s}^{hh}(E)
\end{bmatrix},
\end{equation}
where the submatrix $\tb{s}^{\alpha\beta}(E)$ contains the scattering amplitudes of the processes where a $\beta$-type carrier with energy $E$ injected from the probe is reflected as an $\alpha$-type carrier. For later use, we define $\tb{T}^{\alpha\beta}(E)\equiv\tb{s}^{\alpha\beta\dag}(E)\tb{s}^{\alpha\beta}(E)$. The explicit expression of the scattering matrix of our effective model is presented in Appendix~\ref{appa}.

To set the stage, we briefly review the dc transport properties of the effective model. As analyzed in Ref.~\cite{cao2023differential}, at zero temperature, the differential conductance $G$ and differential shot noise $P$ are given by
\begin{eqnarray}
&&G(V_\tr{dc})=\frac{2e^2}{h}\sum_{s=\pm}\tau_s(eV_\tr{dc}),\label{GV}\\
&&P(V_\tr{dc}) =\frac{8e^3}{h}\sum_{s=\pm}\tau_s(eV_\tr{dc})[1-\tau_s(eV_\tr{dc})],\label{PV}
\end{eqnarray}
where the transmission coefficients $\tau_{\pm}(E)$ correspond to the eigenvalues of the Hermitian matrix $\tb{T}^{he}(E)$. In what follows, we take  $\varepsilon_B=0$ to focus on zero-energy ABSs and MBSs/QMBSs. In this case,
\begin{eqnarray}
\tau_{\pm}(E)  =\frac{\Lambda\pm |E|(\Gamma_{1}-\Gamma_{2})\Pi}{2(E^{2}+\Gamma_{1}^{2})(E^{2}+\Gamma_{2}^{2})}, \label{taupm}
\end{eqnarray}
with
\begin{eqnarray}
\hspace{-0.5cm}&&\Lambda=E^{2}(\Gamma_{1}^{2}+\Gamma_{2}^{2}-2\Gamma_{1}\Gamma_{2}\cos^{2}\theta) +2\Gamma_{1}^{2}\Gamma_{2}^{2}\sin^{2}\theta,\\
\hspace{-0.5cm}&&\Pi =\sqrt{E^{2}(\Gamma_{1}^{2}+\Gamma_{2}^{2}-2\Gamma_{1}\Gamma_{2}\cos2\theta) +( \Gamma_{1}\Gamma_{2}\sin2\theta)^{2}}.
\end{eqnarray}

\begin{figure}[t!]
\centering
\includegraphics[width=0.9\columnwidth]{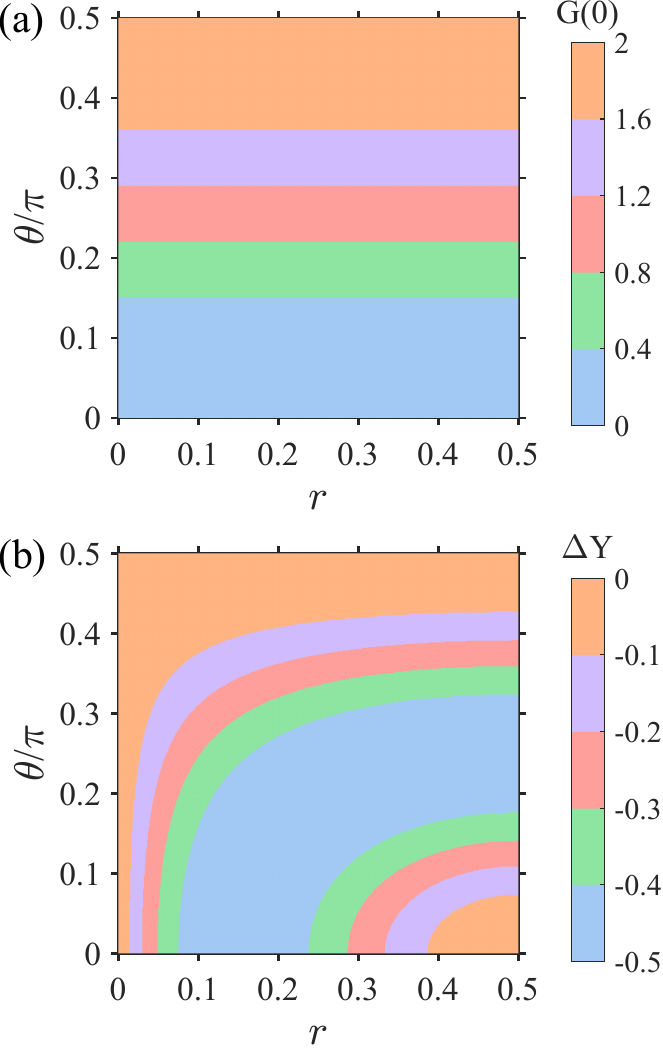}
\caption{(a) Zero-temperature, zero-bias conductance $G(0)$ (in units of $2e^2/h$) and (b) $\Delta Y$, formulated by Eq.~\eqref{deltay2}, as functions of the parameters $r$ and $\theta$ (see Sec.~\ref{secIIA} for their physical meanings). Here $r=0$ corresponds to MBSs/QMBSs, while $r>0$ corresponds to zero-energy ABSs.}\label{Fig2}
\end{figure}

Substituting Eq.~\eqref{taupm} into Eq.~\eqref{GV} yields the zero-temperature, zero-bias conductance,  
\begin{equation}
G(0)=\frac{4e^2}{h}\sin^2\theta. \label{G0}
\end{equation}
The dependence of $G(0)$ on $(r,\theta)$ is shown in Fig.~\ref{Fig2}(a). Qualitatively, ABSs can yield nearly quantized ZBCPs over a broad and continuous parameter range. As shown in Ref.~\cite{cao2023differential}, finite temperature introduces an $r$ dependence in $G(0)$ and enlarges the parameter space supporting nearly quantized ZBCPs. Quantitatively, both MBSs/QMBSs and ABSs with $\theta=0.25\pi$ produce quantized ZBCPs, but they arise from distinct transmission probabilities: $\tau_+(0)=1$ for MBSs/QMBSs, and $\tau_+(0)=\tau_-(0)=1/2$ for ABSs. According to Eq.~\eqref{PV}, $P(V)$ exhibits a zero-bias dip in the former case and a zero-bias peak in the latter, thereby providing a diagnostic to distinguish between MBSs/QMBSs and zero-energy ABSs. However, because the shot noise of a zero-energy state scales with the dc bias amplitude, observing this low-bias feature is experimentally challenging if significant background noise exists in certain setups~\cite{liang2022low,note_noise}.

Now we turn to elucidate how our main findings of excess photon-assisted noise are connected to the scattering matrix. To this end, we define two integrals  
\begin{eqnarray}
&&Y_1 = \sum_{s=\pm} \int_{-\infty}^{\infty} dE \, \tau_{s}^{2}(E), \label{Y1} \\
&&Y_2 = \sum_{s=\pm} \int_{-\infty}^{\infty} dE \, \tau_{s}(E)\,[1-\tau_{s}(E)], \label{Y2}
\end{eqnarray}
and their difference  
\begin{equation}
\Delta Y \equiv Y_1-Y_2. \label{deltay}
\end{equation}
Using the expressions of $\tau_{\pm}(E)$ from Eq.~\eqref{taupm}, one finds $Y_{1} = \pi \Gamma (1-F)^{2}/2$ and $
Y_{2} = \pi \Gamma (1-F^{2})/2$, with $F = 4r(1-r)\cos^{2}\theta$, leading to
\begin{equation}
\Delta Y =-\pi \Gamma F(1-F).\label{deltay2}
\end{equation}
The $(r,\theta)$ dependence of $\Delta Y$ is shown in Fig.~\ref{Fig2}(b). Notably, (i) $\Delta Y=0$ for MBSs/QMBSs and (ii) $\Delta Y\simeq 0$ for zero-energy ABSs in both the upper and lower-right regions. These ABSs can nevertheless be readily distinguished by their zero-bias conductance, which approaches either $0$ or $4e^2/h$, as shown in Fig.~\ref{Fig2}(a). Taken together, Figs.~\ref{Fig2}(a) and \ref{Fig2}(b) demonstrate that $\Delta Y$ provides a criterion to differentiate the ABSs with nearly quantized ZBCPs from MBSs/QMBSs.

In Sec.~\ref{secIID}, we establish an analytical connection between $\Delta Y$ and the excess noise $S^\tr{exc}$, defined in Eq.~\eqref{excessnoise}, at zero temperature. In particular, $\Delta Y=0$ implies that $S^\tr{exc}$ vanishes at nonzero integer values of $eV_\tr{dc}/\Omega$, independent of the functional form of the ac bias. For a harmonic drive $V(t)=V_\tr{dc}[1-\cos(\Omega t)]$, this prediction is confirmed in Sec.~\ref{secIII} by numerical calculations of the full $S^\tr{exc}(eV_\tr{dc}/\Omega)$ dependence, which further highlight a sharp distinction between MBSs/QMBSs and zero-energy ABSs with nearly quantized ZBCPs: while $S^\tr{exc}$ for MBSs/QMBSs undergoes multiple sign reversals as $V_\tr{dc}$ increases, it remains strictly negative for zero-energy ABSs over the entire $V_\tr{dc}$ range.

\subsection{Photon-assisted current and current-noise}\label{secIIC}
We consider a time-dependent bias composed of dc and ac components,  
$V(t)=V_\tr{dc}+V_\tr{ac}(t)$,
where $V_\tr{ac}(t)=V_\tr{ac}(t+\mc T)$ and averages to zero over a period $\mc T$,  
$\frac{1}{\mc T}\int_{0}^{\mc T}dt\,V_\tr{ac}(t)=0$.  
The associated frequency $\Omega=2\pi/\mc T$ sets the photon energy quantum. 

The scattering theory of photon-assisted transport in normal systems was established in Ref.~\cite{pedersen1998scattering}.  
Within this framework, we derive the photon-assisted current and current noise for the superconducting model introduced in Sec.~\ref{secIIA}, as detailed in Appendices~\ref{appb} and \ref{appc}. The two experimental observables, period-averaged current $J$ and zero-frequency current noise $S$, are formally expressed as
\begin{eqnarray}
J &=& \frac{e}{h}\sum_{m}\vert p_{m,e}\vert^{2} \int dE \, 
\Tr[\tb{T}^{he}(E)] \notag\\
&&\times \big[ f_{e}(E-m\Omega) - f_{h}(E+m\Omega) \big], \label{current4}
\end{eqnarray}
and $S=S_1+S_2$, with
\begin{eqnarray}
S_{1} &=& \frac{2e^2}{h} \textrm{Re} \sum_{mm^\prime n}   p_{n,e}^{\ast}p_{m,e}p_{m+m^\prime,e}^{\ast}p_{n+m^\prime,e} \int dE\notag\\
&&\times   K_{1,m^\prime}(E)\Big\{ f_{e}(E-n\Omega)\,[1-f_{e}(E-m\Omega)] \notag\\
&&+ f_{e}(E-m\Omega)\,[1-f_{e}(E-n\Omega)] \Big\}, \label{S1main}\\
S_{2} &=& \frac{2e^2}{h} \textrm{Re} \sum_{mm^\prime n}  p_{n,e}^{\ast}p_{m,h}p_{m+m^\prime,h}^{\ast}p_{n+m^\prime,e}\int dE \notag\\
&&\times   K_{2,m^\prime}(E)\Big\{ f_{e}(E-n\Omega)\,[1-f_{h}(E-m\Omega)] \notag\\
&&+ f_{h}(E-m\Omega)\,[1-f_{e}(E-n\Omega)] \Big\}, \label{S2main}
\end{eqnarray}
where 
\begin{eqnarray}
K_{1,m^\prime}&=&\Tr[\tb{T}^{he}(E)\tb{T}^{he}(E+m^\prime\Omega)],\label{P1}\\
K_{2,m^\prime}&=&\Tr[\tb{s}^{he\dag}(E)\tb{s}^{hh}(E)\tb{s}^{hh\dag}(E+m^\prime\Omega)\tb{s}^{he}(E+m^\prime\Omega)],\label{P2}\notag\\
\end{eqnarray}
and $f_\alpha(E) = 1/\big[e^{(E-\sgn(\alpha)eV_\tr{dc})/k_B T}+1\big]$ is the Fermi distribution of $\alpha$-type carriers in the reservoir. Photon-assisted transport processes manifest themselves through the appearance of $\Omega$ in the energy arguments.

\begin{figure}[t!]
\centering
\includegraphics[width=\columnwidth]{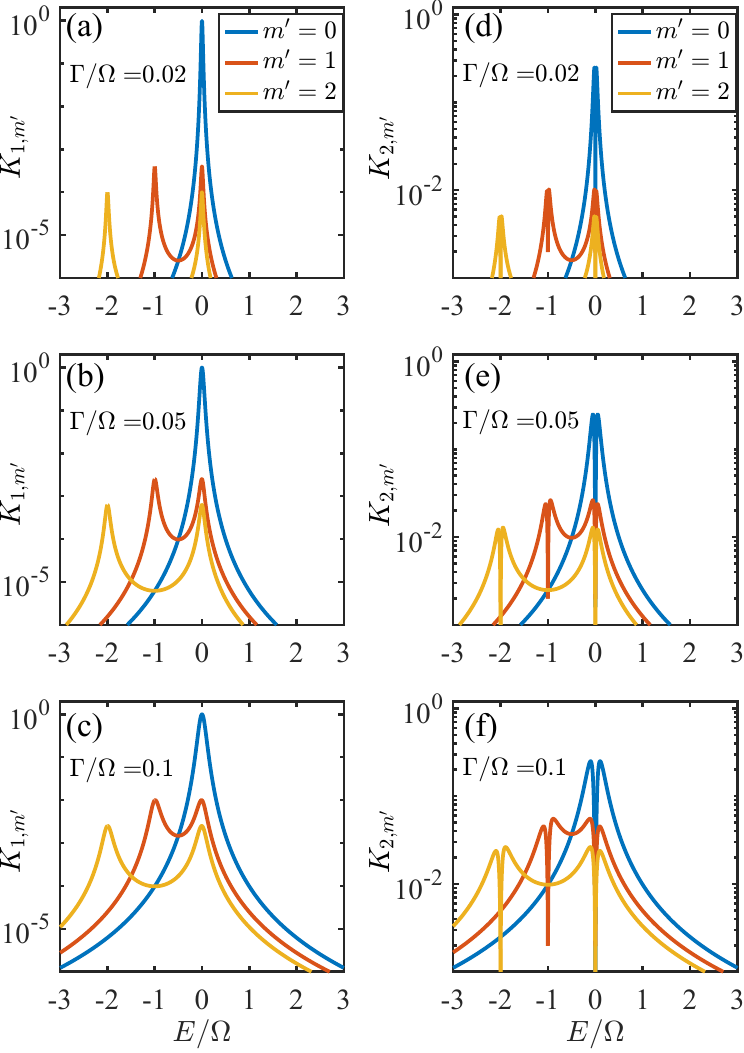}
\caption{[(a)--(c)] $K_{1,m^\prime}$ and [(d)--(f)] $K_{2,m^\prime}$, defined in Eqs.~\eqref{P1} and \eqref{P2}, respectively, as a function of the reduced energy $E/\Omega$ for various $m^\prime$ and $\Gamma/\Omega$ values as indicated. Calculations are performed with $r=0$.}\label{Fig3}
\end{figure}

Since $\Tr[\tb{T}^{he}(E)] = \tau_+(E) + \tau_-(E)$, the current $J$ does not provide useful information beyond the differential conductance of Eq.~\eqref{GV} for distinguishing ABSs from MBSs/QMBSs. In contrast, the explicit expressions for $K_{1,m^\prime}$ and $K_{2,m^\prime}$ indicate that the current noise $S$ encodes higher-order tunneling processes. For consistency, we have verified in Appendix~\ref{appd} that the excess noise in the subgap regime of a normal metal–superconductor tunnel junction, previously obtained using the Keldysh Green's function approach~\cite{belzig2016elementary,bertin2022microscopic}, is exactly reproduced by our scattering-matrix formulation.

\subsection{Analytical insights}\label{secIID}
We are now in a position to demonstrate that the quantity $\Delta Y$, defined in Eq.~\eqref{deltay}, is accessible from the excess noise $S^\tr{exc}$ at nonzero integer values of $eV_\tr{dc}/\Omega$. In Fig.~\ref{Fig3}, we show a set of curves of the functions $K_{1,m^\prime}$ and $K_{2,m^\prime}$ appearing in Eqs.~\eqref{S1main} and \eqref{S2main} for various values of $m^\prime$ and $\Gamma/\Omega$. These curves exhibit characteristic peaks around $E/\Omega=-m^\prime$, with the peak heights decreasing as $|m^\prime|$ increases. As the reduced tunneling strength $\Gamma/\Omega$ increases, the peaks broaden and the peak heights for $m^\prime \neq 0$ tend to approach those for $m^\prime = 0$. Consequently, for $\Gamma/\Omega \ll 1$, it is reasonable to retain only the $m^\prime = 0$ contribution in Eqs.~\eqref{S1main} and \eqref{S2main}. At zero temperature, this yields
\begin{eqnarray}
&&S_1 = \frac{2e^{2}}{h}\sum_{mn} X_{m,n} \bigg\vert \sum_{s=\pm} \int_{(m+v)\Omega}^{(n+v)\Omega} dE \, \tau_s^2(E) \bigg\vert, \label{weakS1} \\
&&S_2 = \frac{2e^{2}}{h}\sum_{mn} X_{m,n} \bigg\vert \sum_{s=\pm} \int_{-(m+v)\Omega}^{(n+v)\Omega} dE \, \tau_s(E) [1-\tau_s(E)] \bigg\vert, \notag\\
\label{weakS2}
\end{eqnarray}
where $v = eV_\tr{dc}/\Omega$ and $X_{m,n} = |p_{m,e} p_{n,e}|^2$. Importantly, when $v$ is an integer $q$, it can be eliminated from the integration bounds via the substitutions $m \to m-q$ and $n \to n-q$. Furthermore, for $\Gamma/\Omega\ll 1$, the integrands in Eqs.~\eqref{weakS1} and \eqref{weakS2} decay rapidly with $|E|$, allowing the integration bounds to be safely extended to $\pm \infty$ or $0$. As a result, the integrals defined in Eqs.~\eqref{Y1} and \eqref{Y2} enter the expressions for $S_1$ and $S_2$, respectively.

Analytically, for $eV_\tr{dc}/\Omega=q \ne 0$, we obtain (see Appendix~\ref{appe} for details)  
\begin{eqnarray}
S^\tr{exc} = \frac{2e^{2}}{h} \Big( \kappa \, \Delta Y - \vert p_{-q,e}\vert^4 \, \overline{Y} \Big), \label{weakdS}
\end{eqnarray}
where $\kappa = \sum_{mn<0} X_{m-q,n-q} + \tfrac{1}{2} \sum_{mn=0} X_{m-q,n-q}$ and $\overline{Y}=(Y_1+Y_2)/2$.  
Evidently, $S^\tr{exc}$ scales linearly with $\Delta Y$ up to a small fourth-order correction, offering a direct experimental test of whether $\Delta Y=0$.  

Notably, Eq.~\eqref{weakdS} is valid irrespective of the functional form of the ac drive $V_\tr{ac}(t)$. In experiments on excess noise in normal tunnel junctions~\cite{dubois2013minimal,gabelli2013shaping}, harmonic, square, and Lorentzian ac biases have been employed. In the following, we focus on a harmonic bias, $V(t)=V_\tr{dc}[1-\cos(\Omega t)]$, due to its simplicity and experimental accessibility. For this choice, $p_{m,e}=J_{m}(-v)$ and $p_{m,h}=J_{-m}^\ast(-v)$, where $J_{m}$ is the Bessel function of the first kind of order $m$.

\begin{figure}[t!]
\centering
\includegraphics[width=0.9\columnwidth]{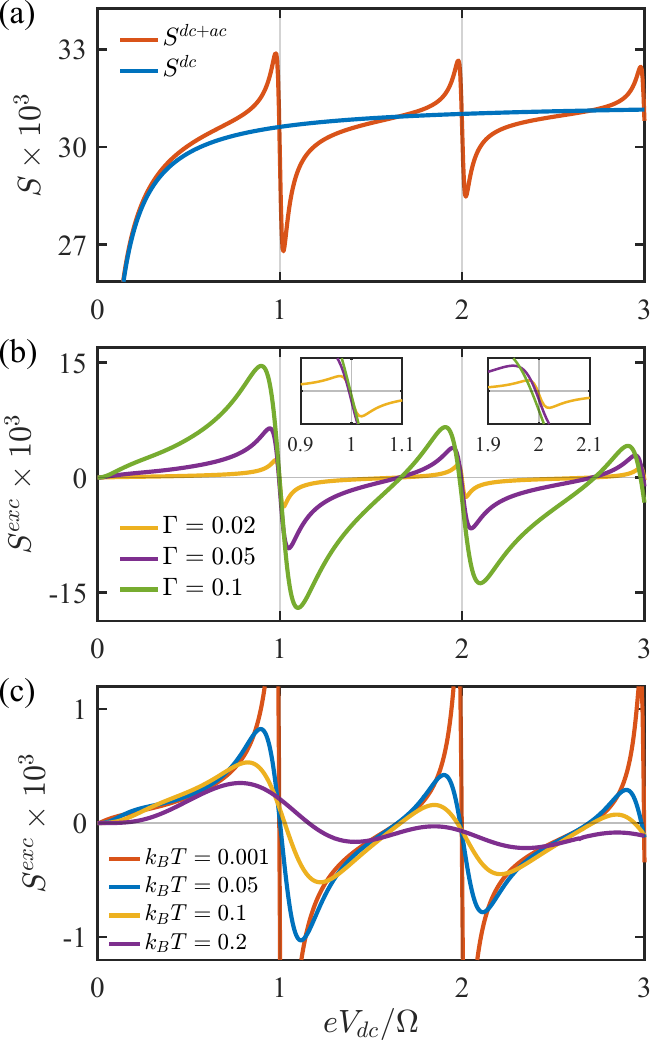}
\caption{Current noise as a function of the reduced dc bias $eV_{\mathrm{dc}}/\Omega$ for MBSs/QMBSs. 
Here, $S^{\mathrm{dc}}$ and $S^{\mathrm{dc+ac}}$ denote the current noise with the ac bias component switched off and on, respectively, while $S^{\mathrm{exc}}$ represents the excess noise defined as $S^{\mathrm{dc+ac}}-S^{\mathrm{dc}}$. 
The insets in (b) show magnified views around $eV_{\mathrm{dc}}/\Omega = 1$ and $2$. 
The parameters are: (a) $k_{B}T=0.001$ and $\Gamma=0.02$; (b) $k_{B}T=0.001$; and (c) $\Gamma=0.02$. 
Other parameters are specified in each panel.}\label{Fig4}
\end{figure}

\begin{figure}[t!]
\centering
\includegraphics[width=0.9\columnwidth]{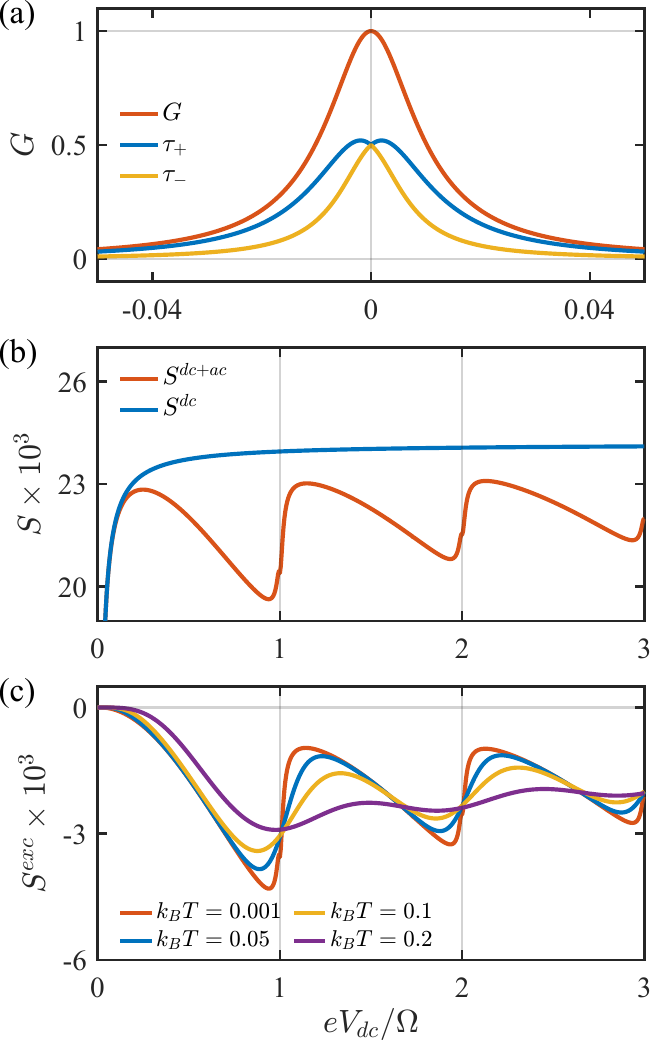}
\caption{Conductance and current noise for a zero-energy ABS with $r=0.4$ and $\theta=0.25\pi$. (a) The ABS exhibits a quantized ZBCP arising from the additive contributions of two eigenchannels with transmissions $\tau_+$ and $\tau_-$. (b) Current noise as a function of the reduced dc bias $eV_\tr{dc}/\Omega$ at $k_BT=0.001$ and $\Gamma=0.02$, where $S^\tr{dc}$ and $S^\tr{dc+ac}$ denote the noise with the ac bias component switched off and on, respectively. (c) Excess noise for $\Gamma=0.02$ at different temperatures as indicated.}\label{Fig5}
\end{figure}

\section{Numerical results and discussions}\label{secIII}
\subsection{Effective model calculations}\label{secIIIA}
The analytical results above are limited to the regime where $eV_\tr{dc}/\Omega$ takes nonzero integer values, $\Gamma/\Omega \ll 1$, and $k_B T=0$.  
To go beyond these restrictions, we numerically evaluate the current noise using Eqs.~\eqref{S1main} and \eqref{S2main}.  
In this section, all energies $eV_\tr{dc}$, $\Gamma$, and $k_B T$ are expressed in units of $\Omega$, while the conductance $G$ and current noise $S$ shown in the figures are given in units of $2e^2/h$ and $2e^2\Omega/h$, respectively.

Figure~\ref{Fig4} shows the current noises for MBSs/QMBSs. In Fig.~\ref{Fig4}(a), we present the noises as a function of the reduced dc bias $eV_\tr{dc}/\Omega$ at $k_BT=0.001$ and $\Gamma=0.02$. Here, $S^\tr{dc}$ and $S^\tr{dc+ac}$ denote the noise with the ac bias component switched off and on, respectively. As the dc bias increases, $S^\tr{dc}$ rises monotonically and eventually saturates, whereas $S^\tr{dc+ac}$ exhibits antisymmetric line shapes around $eV_\tr{dc}/\Omega=q$, with $q$ being nonzero integers. Notably, $S^\tr{dc}$ and $S^\tr{dc+ac}$ intersect exactly at $eV_\tr{dc}/\Omega=q$, leading to vanishing excess noise in agreement with our analytical results. This feature is more clearly highlighted in Fig.~\ref{Fig4}(b), where the excess noise $S^\tr{exc}$ undergoes multiple sign reversals. When $\Gamma$ increases, the overall amplitude of the $S^\tr{exc}$ curve grows. Meanwhile, as shown in the insets of Fig.~\ref{Fig4}(b), $S^\tr{exc}(eV_\tr{dc}/\Omega=1)$ remains pinned at zero with increasing $\Gamma$, while $S^\tr{exc}(eV_\tr{dc}/\Omega=2)$ deviates from zero, reflecting the enhanced contributions of the $m^\prime \neq 0$ terms in Eqs.~\eqref{S1main} and \eqref{S2main} to current noise at larger bias. Figure~\ref{Fig4}(c) demonstrates that the feature $S^\tr{exc}=0$ at $eV_\tr{dc}/\Omega=q$ also breaks down once thermal energy becomes much larger than $\Gamma$. In this regime, the $S^\tr{exc}$ curve is broadened and its amplitude reduced. 

\begin{figure}[t!]
\centering
\includegraphics[width=\columnwidth]{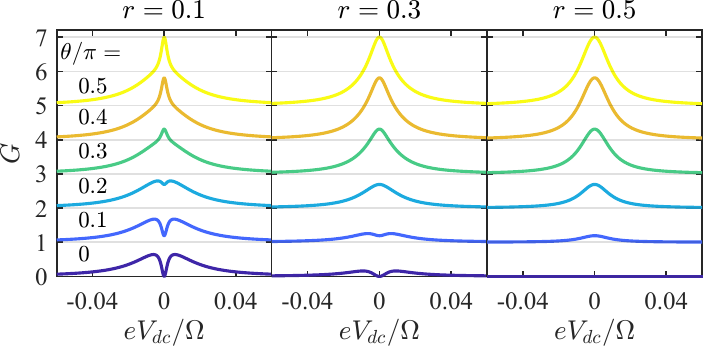}
\caption{Zero-temperature differential conductance of various zero-energy ABSs characterized by different $(r,\theta)$ combinations at $\Gamma=0.02$. 
Similar to Fig.~\ref{Fig5}(a), the conductance arises from the additive contributions of two eigenchannels with transmissions $\tau_+$ and $\tau_-$. 
The conductance curve exhibits either a zero-bias peak or a zero-bias dip; see text for their physical origins.
}\label{Fig6}
\end{figure}

\begin{figure}[htbp]
\centering
\includegraphics[width=\columnwidth]{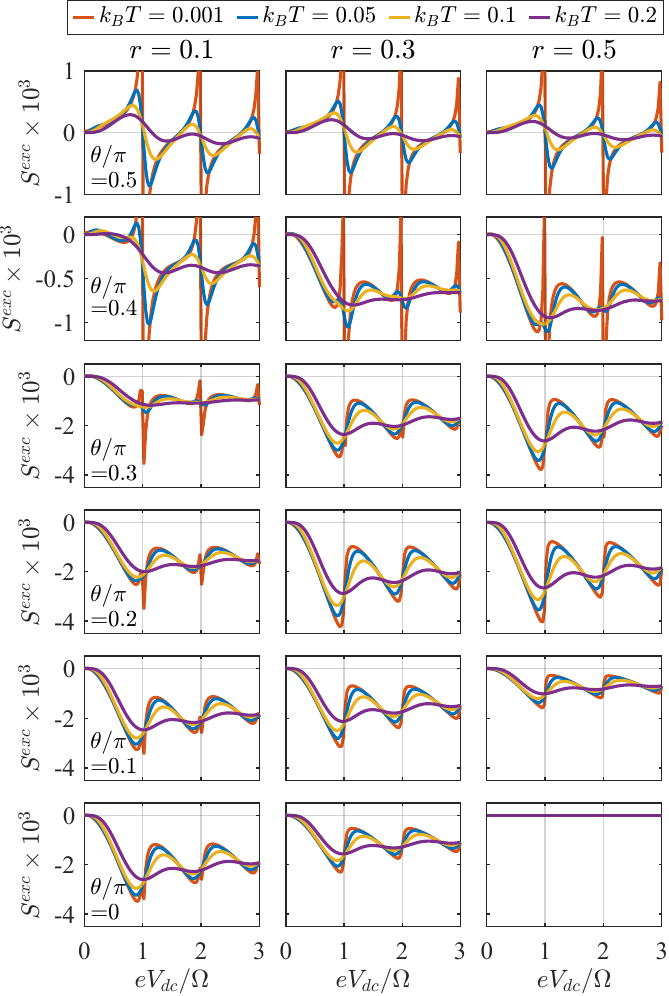}
\caption{Excess noise $S^\tr{exc}$ of zero-energy ABSs for different $(r,\theta)$ combinations at $\Gamma=0.02$ and various temperatures. 
These panels correspond one-to-one with those in Fig.~\ref{Fig6}. For most ABSs, except those with $\theta/\pi=0.5$, the feature of $S^\tr{exc}=0$ predicted for MBSs/QMBSs is absent when $k_B T \gtrsim \Gamma$.}\label{Fig7}
\end{figure}

For comparison, Fig.~\ref{Fig5} presents the excess noise for a zero-energy ABS characterized by $r=0.4$ and $\theta=0.25 \pi$. As shown in Fig.~\ref{Fig5}(a), this state produces a quantized ZBCP originating from two eigenchannels with transmission coefficients $\tau_+$ and $\tau_-$. In Fig.~\ref{Fig5}(b), $S^\tr{dc+ac}$ displays a sawtooth line shape and remains below $S^\tr{dc}$ over the entire dc bias range, in sharp contrast to the noise behaviors in Fig.~\ref{Fig4}(a). In Fig.~\ref{Fig5}(c), the temperature effect on the excess noise $S^\tr{exc}$ resembles that of the MBSs/QMBSs in Fig.~\ref{Fig4}(c).  

For completeness, we present the differential conductance $G$ and excess noise $S^\tr{exc}$ for various zero-energy ABSs by sweeping the $(r,\theta)$ parameter space. Figure \ref{Fig6} shows the zero-temperature conductance, where each column corresponds to a fixed $r$ value with varying $\theta$, as indicated in the first column. Similar to Fig.~\ref{Fig5}(a), these conductance curves arise from the additive contributions of two eigenchannels with transmissions $\tau_+$ and $\tau_-$. The $G(0)$ values are independent of $r$ and equal $4e^2\sin^2\theta/h$ as predicted by Eq.~\eqref{G0}. Moreover, the conductance line shapes can be grouped into three categories: a zero-bias peak larger than $2e^2/h$, a zero-bias peak smaller than $2e^2/h$, and a zero-bias dip. This can be understood as follows. As illustrated in Fig.~\ref{Fig1}(a), Andreev tunneling occurs between the probe and the two Majorana components $\gamma_1$ and $\gamma_2$ of an ABS. The zero-bias peaks (dips) in Fig.~\ref{Fig6} originate from constructive (destructive) interference between these two processes. Specifically, for $\theta=0$, where $\lambda_{1\uparrow}=\lambda_{2\uparrow}=0$, both $\gamma_1$ and $\gamma_2$ couple to the spin-down channel of the probe, and their destructive interference yields $G(0)=0$. For $\theta=0.5\pi$, where $\lambda_{1\uparrow}=\lambda_{2\downarrow}=0$, $\gamma_1$ and $\gamma_2$ couple to opposite spin channels, and the associated resonant Andreev reflections add constructively, resulting in a total conductance $G(0)=4e^2/h$.  

Figure~\ref{Fig7} shows the excess noise $S^\tr{exc}$ of zero-energy ABSs characterized by different $(r,\theta)$ combinations at various temperatures. The panels correspond one-to-one to those in Fig.~\ref{Fig6}. For $\theta \leq 0.3\pi$, the $S^\tr{exc}$ curves remain negative over the entire dc bias range. As temperature increases, the local maxima and minima of $S^\tr{exc}$ shift downward and upward, respectively. For $\theta = 0.4\pi$, the $S^\tr{exc}$ curves at low temperatures ($k_B T<\Gamma$) display sharp variations around $eV_\tr{dc}/\Omega=q$, producing $S^\tr{exc} \approx 0$ at these points. Nevertheless, this behavior is suppressed as the temperature increases. For $\theta=0.5\pi$, the $S^\tr{exc}$ curves closely reproduce those of MBSs/QMBSs shown in Fig.~\ref{Fig4}(c). This is because $\gamma_1$ and $\gamma_2$ couple to opposite spin channels of the probe, as discussed earlier. These two independent resonant Andreev reflection processes therefore produce noise features identical to those of a single MBSs/QMBSs.  

These results indicate that the combined analysis of differential conductance and excess noise provides a practical means to distinguishing zero-energy ABSs, especially those producing nearly quantized ZBCPs, from MBSs/QMBSs.

\subsection{Simulations of hybrid nanowires}\label{secIIIB}
The results in the preceding section are obtained from a platform-independent effective model. To verify their validity, we calculate the excess noise of MBSs and zero-energy ABSs in semiconductor–superconductor hybrid nanowires, which constitute one of the most promising platforms for topological quantum computation~\cite{plugge2017majorana,karzig2017scalable}. The hybrid nanowire can be modeled by the one-dimensional Bogoliubov–de Gennes (BdG) Hamiltonian~\cite{lutchyn2010majorana,oreg2010helical}
$H_\textrm{wire}=\int_{0}^{L} dx\, \Psi^{\dag}(x){\cal H}\Psi(x)$,
where $\Psi^{\dag}=(\psi_\uparrow^\dag,\psi_\downarrow^\dag,\psi_\downarrow,-\psi_\uparrow)$ and
${\cal H}=\left[ -\frac{\partial_{x}^{2}}{2m^{\ast}}-\mu+V_\textrm{pot}(x)+i\alpha\sigma_{y}\partial_x \right]\tau_{z}+V_{Z}\sigma_{x}+\Delta(x)\tau_{x}$.
Here $L$, $m^\ast$, $\mu$, $V_\textrm{pot}$, $\alpha$, $V_Z$, and $\Delta$ denote the wire length, effective electron mass, chemical potential, electrostatic potential, Rashba spin–orbit coupling, Zeeman energy, and proximity-induced superconducting pairing potential, respectively, in the semiconductor. The Pauli matrices $\sigma_{x,y,z}$ ($\tau_{x,y,z}$) act in spin (electron–hole) space. In realistic tunneling spectroscopy experiments, a short portion of the wire adjacent to the electrode serves as the tunnel junction controlled by a nearby gate. Consequently, the electrostatic potential and superconducting pairing potential near the junction region may differ from those farther away, depending sensitively on device details. These nonuniformities are captured by the position dependence of $\Delta(x)$ and $V_\textrm{pot}(x)$~\cite{liu2017Andreev,fleckenstein2018decaying}.

In numerical simulations, we use the parameters $m^\ast=0.026\,m_e$, $\alpha=30\,\text{meV\,nm}$, and $L=2\,\mu$m, typical for InAs–Al hybrid nanowires~\cite{lutchyn2018majorana}. For an isolated nanowire, diagonalization of $H_\textrm{wire}$ on a lattice with spacing $a=10$ nm yields the energy spectrum and wave functions. When the left end of the wire is attached to an electrode, we compute the scattering matrix to evaluate the differential conductance and current noise, as outlined in Sec.~\ref{secII}. We set the hopping amplitude between the electrode and the leftmost lattice site of the wire to $\lambda=\hbar^2/2m^\ast a^2$, with electrode density of states $\rho=0.01/\lambda$. As will be shown later, these parameters yield ZBCP widths of $\Gamma\simeq0.01$ meV. To satisfy the condition $\Gamma/\Omega \ll 1$, we take $\Omega=0.05$ meV ($\sim$ 12 GHz), within the experimental range in photon-assisted noise measurements~\cite{schoelkopf1998observation,reydellet2003quantum,gasse2013observation,dubois2013minimal,gabelli2013shaping,kozhevnikov2000observation}. The dc bias amplitude is further restricted to values well below the superconducting gap, ensuring that quasiparticles outside the gap can be neglected and the formulas in Sec.~\ref{secIIC} remain valid. Notably, in recently developed PbTe–Pb hybrid nanowires~\cite{cao2022numerical,jiang2022selective,zhang2023proximity,gao2025quantized}, the induced gap can be several times larger than that of InAs-Al since Pb has a much larger superconducting gap than Al, enabling noise measurements in the deep Andreev tunneling regime.

\begin{figure}[t!]
\centering
\includegraphics[width=0.9\columnwidth]{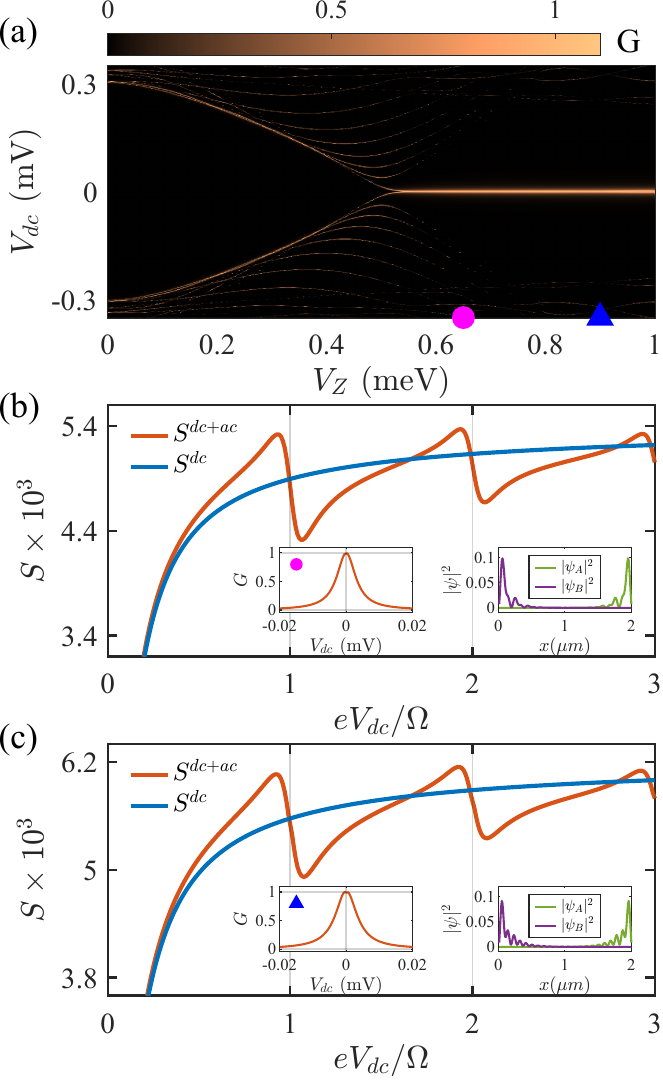}
\caption{Simulation results for a uniform hybrid nanowire. (a) Differential conductance map as a function of Zeeman energy $V_Z$ and dc bias $V_\tr{dc}$. [(b), (c)] For the two zero-energy states appearing at different $V_Z$ values indicated by the markers in (a), we show their current noise versus $eV_\tr{dc}/\Omega$ in the main panels, differential conductance line cuts from (a) in the left insets, and spatial profiles of the wave functions in the Majorana basis in the right insets. Here, $\Omega = 0.05$ meV, and the current noise is given in units of $2e^3,\tr{mV}/h$.}\label{Fig8}
\end{figure}

For a uniform hybrid nanowire with $\mu=0.5$ meV, $V_\textrm{pot}(x)=0$, and $\Delta(x)=0.3$ meV, the differential conductance map versus $V_Z$ and $V_\tr{dc}$ is shown in Fig.~\ref{Fig8}(a). ZBCPs emerge once $V_Z\gtrsim0.54$ meV, close to the critical field $V_Z^C=\sqrt{\Delta^2+\mu^2}\approx0.58$ meV expected for an isolated, infinitely long wire at the topological superconducting phase transition~\cite{lutchyn2010majorana,oreg2010helical}. The vertical cuts at the two markers exhibit quantized ZBCPs, see left insets of Figs.~\ref{Fig8}(b) and \ref{Fig8}(c). The nature of the bound states responsible for the quantized ZBCPs can be identified by examining the corresponding spatial wave functions in the Majorana basis. To be more precise, the eigenenergies obtained from diagonalizing $H_\textrm{wire}$ always come in pairs as $\pm E_n$, due to the intrinsic particle–hole symmetry of a BdG Hamiltonian. Denoting the wave functions of the lowest positive- and negative-energy eigenstates as $\psi_{\pm}(x)$, we construct the wave functions in the Majorana basis as~\cite{beenakker2015random,moore2018two}:
$\psi_{A}(x)=\tfrac{1}{\sqrt{2}}\,[\psi_{+}(x)+\psi_{-}(x)]$,
$\psi_{B}(x)=\tfrac{i}{\sqrt{2}}\,[\psi_{+}(x)-\psi_{-}(x)]$.
As shown in the right insets of Figs.~\ref{Fig8}(b) and \ref{Fig8}(c), these wave functions localize at opposite ends of the wire, confirming that the quantized ZBCPs indeed originate from MBSs. In the main panels of Figs.~\ref{Fig8}(b) and \ref{Fig8}(c), $S^\tr{dc}$ and $S^\tr{dc+ac}$ curves intersect at nonzero integer values of $eV_\textrm{dc}/\Omega$, leading to vanishing excess noise, in agreement with our effective model predictions for MBSs.  

\begin{figure}[t!]
\centering
\includegraphics[width=0.9\columnwidth]{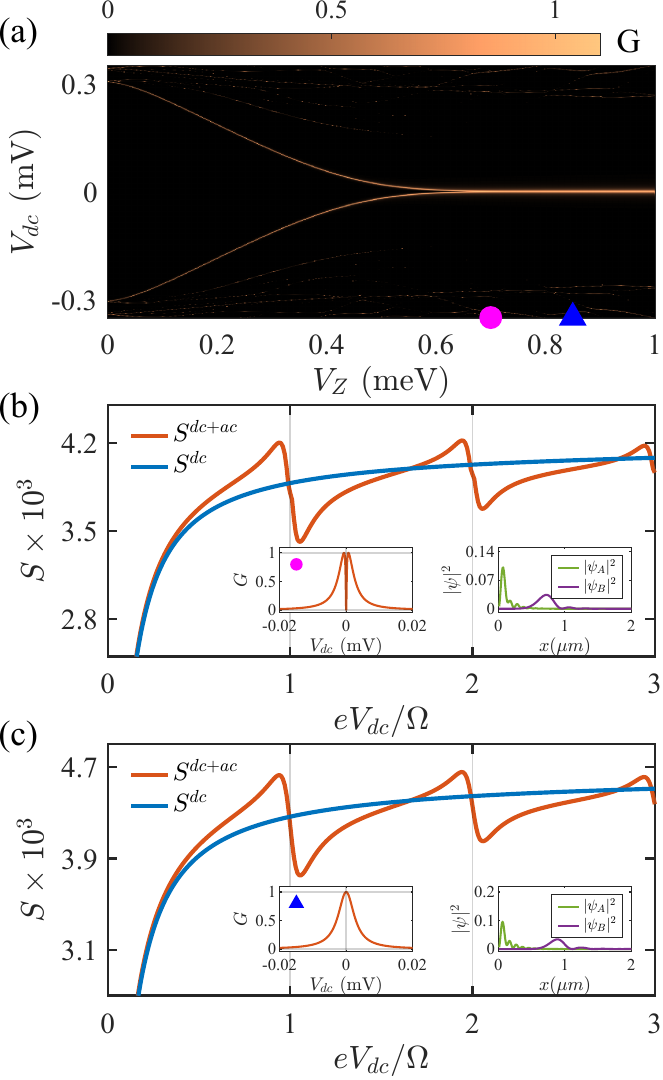}
\caption{Similar to Fig.~\ref{Fig8}, but for a hybrid nanowire with a smoothly varying potential near the left end. In panels (b) and (c), the wave functions shown in the right insets are spatially separated, with $\psi_B$ located in the interior of the wire, indicating that the ZBCPs shown in the left insets arise from QMBSs rather than MBSs. Nevertheless, the current noise curves behave similarly to those in Fig.~\ref{Fig8}.
}\label{Fig9}
\end{figure}

\begin{figure}[t!]
\centering
\includegraphics[width=0.9\columnwidth]{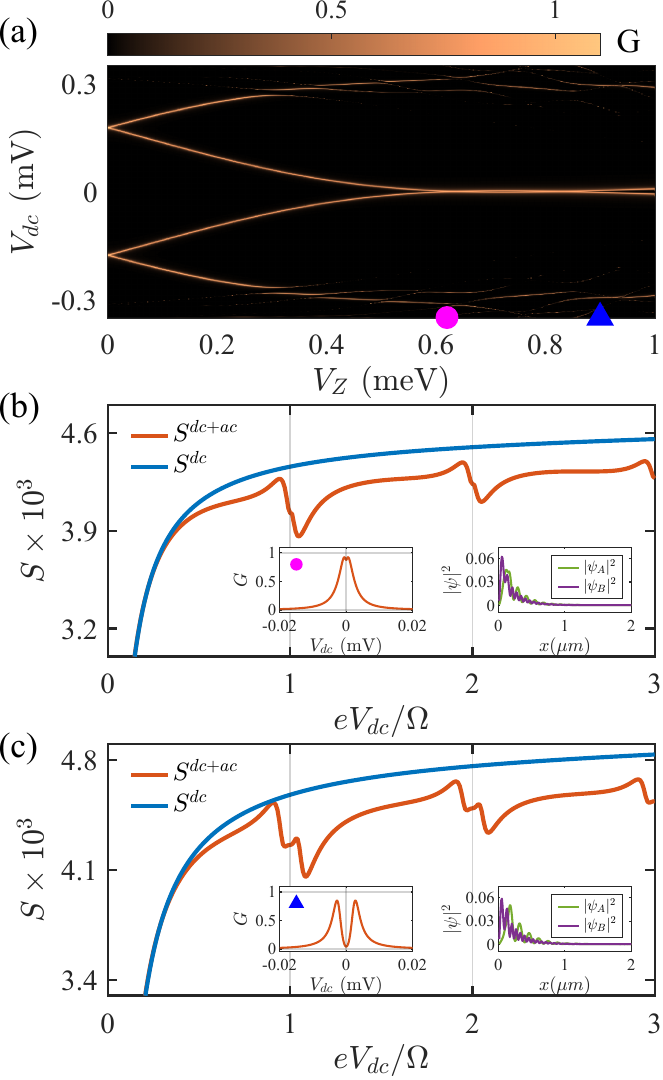}
\caption{Similar to Fig.~\ref{Fig8}, but for a hybrid nanowire with a quantum dot at the left end. In panels (b) and (c), the wave functions shown in the right insets are strongly overlapped, indicating that the ZBCPs shown in the left insets arise from ABSs rather than MBSs. 
The $S^{\mathrm{dc+ac}}$ curves do not surpass the $S^{\mathrm{dc}}$ curves, in contrast to those shown in Fig.~\ref{Fig8}.
}\label{Fig10}
\end{figure}

It is well established that QMBSs can emerge when a smoothly varying electrostatic potential is present near the tunnel-junction region~\cite{vuik2019reproducing,moore2018two}. To model this situation, we set $\mu = 1$ meV, $\Delta(x) = 0.3$ meV, and $V_{\mathrm{pot}}(x) = \mu (1 - x)\,\Theta(1 - x)$. The QMBS nature of the resulting low-energy state is directly reflected in the wave functions $\psi_A$ and $\psi_B$ shown in the right insets of Figs.~\ref{Fig9}(b) and \ref{Fig9}(c). Notably, the conductance map and current noise curves in Fig.~\ref{Fig9} display the same qualitative features as the MBS results in Fig.~\ref{Fig8}. The underlying reason is that, in both the MBS and QMBS cases, the component $\psi_B$ has negligible weight at the left end of the wire, so that the local transport is determined exclusively by $\psi_A$. This fact is captured by our effective model, in which both MBSs and QMBSs are characterized by the parameter value $r = 0$.

We now consider a hybrid nanowire with a quantum dot at the left end, as reported experimentally~\cite{deng2016majorana}. We set the dot length $l_d$ to $0.15~\mu$m and use the parameters $\mu=1.8$ meV, $V_\textrm{pot}(x)=1.17\,\Theta(0.15-l_d)$ meV, and $\Delta(x)=0.3\,\Theta(x-l_d)$ meV, where $\Theta(x)$ is the Heaviside step function. The differential conductance map versus $V_Z$ and $V_\tr{dc}$ is shown in Fig.~\ref{Fig10}(a). Clearly, at $V_Z=0$, there exist two finite-energy ABSs within the gap. As $V_Z$ increases, they split into four branches. Around $V_Z\simeq 0.6$ meV, the two inner branches merge into a zero-energy state, which persists as $V_Z$ increases and exhibits a small splitting when $V_Z\gtrsim 0.9$ meV. The vertical cuts at the two markers highlight nearly quantized ZBCPs, as shown in the left insets of Figs.~\ref{Fig10}(b) and \ref{Fig10}(c). We note that the nearly quantized ZBCPs emerging for $0.6\,\tr{meV}\lesssim V_Z\lesssim0.9$ meV result from a fine-tuning of $V_\textrm{pot}(x)$. The corresponding Majorana wave functions in the right insets of Figs.~\ref{Fig10}(b) and \ref{Fig10}(c) are localized at the same end of the wire, confirming that these ZBCPs arise from ABSs rather than MBSs/QMBSs. In this case, $S^\tr{dc+ac}$ remains below $S^\tr{dc}$ over the entire bias range, again consistent with our effective model predictions for zero-energy ABSs.  

\section{Summary and discussions}\label{secIV}
We have investigated photon-assisted shot noise in tunneling into zero-energy ABSs or MBSs/QMBSs under a combined dc and ac bias, providing insights that complement earlier theoretical studies~\cite{golub2011shot,perrin2022identifying,cao2023differential,wong2022shot} restricted to dc bias. The ac drive enables photon-assisted transport processes, with frequency $\Omega$ (in the GHz range) setting the photon energy. We assume the energy hierarchy $k_{B}T \lesssim \Gamma \ll \Omega$ and focus on the regime where Andreev tunneling dominates while quasiparticle tunneling is suppressed. Within an effective tunneling model, we analytically demonstrate that the excess photon-assisted noise $S^{\mathrm{exc}}$ of MBSs/QMBSs vanishes at nonzero integer values of $eV_{\mathrm{dc}}/\Omega$, independent of the specific form of the ac drive. For a harmonic bias $V(t)=V_{\mathrm{dc}}[1-\cos(\Omega t)]$, numerical calculations confirm this prediction and further reveal that $S^{\mathrm{exc}}$ exhibits multiple sign reversals as $V_{\mathrm{dc}}$ increases. By contrast, for zero-energy ABSs—particularly those producing nearly quantized ZBCPs—$S^{\mathrm{exc}}$ remains strictly negative over the entire $V_\tr{dc}$ range. These results are further corroborated by numerical simulations of semiconductor–superconductor hybrid nanowires hosting either MBSs or zero-energy ABSs.

Since our predictions for excess photon-assisted noise rely solely on a single local probe, they can, in principle, be tested on any platform where tunneling conductance has been measured using an electrode or an STM tip. We emphasize, however, that local measurements of tunneling conductance and excess photon-assisted noise cannot unambiguously distinguish QMBSs from MBSs, particularly when QMBSs exhibit robustness against variations in experimental parameters. Nevertheless, such measurements provide a practical means of discriminating trivial zero-energy ABSs—typically arising from disorder or spatial inhomogeneity—from MBSs/QMBSs. This capability enables a meaningful assessment of platform quality by ruling out zero-energy ABSs. In sufficiently clean systems with advanced nanofabrication, unambiguous identification of MBSs can then be pursued through complementary measurement schemes in more sophisticated device architectures~\cite{pikulin2021protocol,bolech2007observing,nilsson2008splitting,akhmerov2011quantized,zhang2019next,liu2021topological,valentini2022majorana,sbierski2022identifying,sergey2022revealing,cao2023recent,dmytruk2023microwave,ren2024microwave}.

\section*{Acknowledgements}
This work was supported by the National Natural Science Foundation of China (Grants No.~12374158, No.~12074039, and No.~92565302), Beijing Natural Science Foundation (Grant No.~1262049), and the Innovation Program for Quantum Science and Technology (Grant No.~2021ZD0302400).

\section*{Data availability}
The data that support the findings of this article are not publicly available. The data are available from the authors upon reasonable request.

\appendix
\section{Scattering matrix}\label{appa}
The scattering matrix of the effective tunneling model under consideration can be obtained by the Mahaux--Weidenm\"{u}ller formula~\cite{mahaux1969shell}  
\begin{equation}
\tb{s}(E)=
\begin{bmatrix}
\tb{s}^{ee}(E) & \tb{s}^{eh}(E) \\
\tb{s}^{he}(E) & \tb{s}^{hh}(E)
\end{bmatrix}
=\tb{I}-2\pi i\,\tb{W}^{\dag} \tb{G}^{r}(E)\tb{W},\label{mahaux}
\end{equation}
where $\tb{I}$ is the identity matrix, $\tb{G}^{r}(E)$ is the dressed retarded Green’s function of the bound state, and $\tb{W}$ describes the coupling between the probe and the bound state. Explicitly, they are given by
\begin{eqnarray}
&&\tb{G}^{r}(E)=\big(E-\tb{H}_{B}+i\pi \tb{W}\tb{W}^{\dag}\big)^{-1},\\
&&\tb{H}_{B}=
\begin{bmatrix}
0 & i\varepsilon_{B}\\
-i\varepsilon_{B} & 0
\end{bmatrix},\\
&&\tb{W}=\sqrt{\rho}
\begin{bmatrix}
\lambda_{1\uparrow}^{\ast} & \lambda_{1\downarrow}^{\ast} & -\lambda_{1\uparrow} & -\lambda_{1\downarrow}\\
\lambda_{2\uparrow}^{\ast} & \lambda_{2\downarrow}^{\ast} & -\lambda_{2\uparrow} & -\lambda_{2\downarrow}
\end{bmatrix}.
\end{eqnarray}

The unitarity of the scattering matrix, $\tb{s}^\dag(E)\tb{s}(E)=\tb{I}$, leads to the following relationships for the submatrices in Eq.~\eqref{mahaux}:
\begin{eqnarray}
&&\tb{s}^{\alpha\alpha\dag}(E)\tb{s}^{\alpha\alpha}(E) + \tb{s}^{\bar{\alpha}\alpha\dag}(E)\tb{s}^{\bar{\alpha}\alpha}(E) = \tb{I}, \label{B1}\\
&&\tb{s}^{\alpha\alpha\dag}(E)\tb{s}^{\alpha\bar{\alpha}}(E) = -\tb{s}^{\bar{\alpha}\alpha\dag}(E)\tb{s}^{\bar{\alpha}\bar{\alpha}}(E). \label{B2}
\end{eqnarray}
Similarly, considering $\tb{s}(E)\tb{s}^\dag(E)=\tb{I}$, one obtains
\begin{equation}
\tb{s}^{\alpha\alpha}(E)\tb{s}^{\alpha\alpha\dag}(E) + \tb{s}^{\alpha\bar{\alpha}}(E)\tb{s}^{\alpha\bar{\alpha}\dag}(E) = \tb{I}. \label{B3}
\end{equation}
Combining Eqs.~\eqref{B1} and \eqref{B3} yields the trace identities
\begin{equation}
\Tr\!\left[\tb{I}-\tb{T}^{\alpha\alpha}(E)\right] = \Tr\!\left[\tb{T}^{\bar{\alpha}\alpha}(E)\right] = \Tr\!\left[\tb{T}^{\alpha\bar{\alpha}}(E)\right]. \label{B4}
\end{equation}

In the Nambu basis $\{\hat{a}_{e\uparrow},\hat{a}_{e\downarrow},\hat{a}_{h\uparrow},\hat{a}_{h\downarrow}\}^{T}$, the intrinsic particle–hole symmetry of the BdG Hamiltonian is expressed as $\Xi H_\tr{BdG}=-H_\tr{BdG} \Xi$, with $\Xi=\tau_x K$, where $K$ denotes complex conjugation and $\tau_x$ is the first Pauli matrix in the electron–hole space. Consequently, the scattering matrix satisfies $\Xi \tb{s}(E)=\tb{s}(-E) \Xi$, which implies
\begin{eqnarray}
&&\tb{s}^{\alpha\beta}(E) = \tb{s}^{\bar{\alpha}\bar{\beta}\ast}(-E), \label{B5}\\
&&\tb{T}^{\alpha\beta}(E) = \tb{T}^{\bar{\alpha}\bar{\beta}\ast}(-E). \label{B6}
\end{eqnarray}

These relations are subsequently employed in Appendices \ref{appb} and \ref{appc} to derive the expressions for the current and current noise presented in the main text.

\section{Derivation of period-averaged current}\label{appb}
In the presence of an ac bias, an electron ($e$) or hole ($h$) emitted from the reservoir connected to the probe acquires an additional time-dependent phase while propagating toward the scattering interface between the probe and the bound state. On top of its plane-wave form $\exp[i(kx-Et)]$, the carrier picks up a phase  
\begin{eqnarray}
\Phi_{\alpha}(t)=\exp\!\left[-i\,\sgn(\alpha)\,e\int_{-\infty}^{t}dt^{\prime}\,V_\tr{ac}(t^{\prime})\right], \label{Phit}
\end{eqnarray}
with $\alpha=\{e,h\}$, $\sgn(e)=1$, and $\sgn(h)=-1$. Since $V_\tr{ac}(t)$ averages to zero over one period $\mc T$, $\Phi_\alpha(t)$ is periodic and can be expanded in a Fourier series, $\Phi_{\alpha}(t)=\sum_m p_{m,\alpha} \exp[-im\Omega t]$. The wave function then takes the form $\sum_m p_{m,\alpha} \exp \{i[kx-(E+m\Omega)t]\}$, representing a superposition of plane waves with shifted energies. Each term corresponds to a process in which a carrier with energy $E$ absorbs ($m>0$) or emits ($m<0$) $|m|$ photons, with $p_{m,\alpha}$ denoting the associated probability amplitudes. This leads to a relation 
\begin{equation}
\hat{a}_{\alpha\sigma}(E)=\sum_{m} p_{m,\alpha}\,\hat{a}_{\alpha\sigma}^{\prime}(E-m\Omega), \label{aa'}
\end{equation}
where $\hat{a}_{\alpha\sigma}$ and $\hat{a}_{\alpha\sigma}^{\prime}$ denote annihilation operators for incoming $\alpha$-type carriers with spin $\sigma$ near the scattering interface and in the reservoir, respectively.

Tunneling between the probe and the bound state is governed by the scattering matrix in Eq.~\eqref{mahaux}, which relates incoming and outgoing states,
\begin{equation}
\hat{b}_{\alpha\sigma}(E)=\sum_{\beta\sigma^\prime} 
\tb{s}^{\alpha\sigma,\beta\sigma^\prime}(E)\,\hat{a}_{\beta\sigma^\prime}(E), \label{ba}
\end{equation}
where $\hat{b}_{\alpha\sigma}$ annihilates an outgoing $\alpha$-type carrier with spin $\sigma$ near the scattering interface.  

The difference in occupation number between incoming and outgoing carriers gives the charge current operator
\begin{eqnarray}
\hat{J}(t) &=& \frac{e}{2h}\sum_{\alpha\sigma}\sgn(\alpha)  
\int dE \int dE^{\prime}\, e^{i(E-E^{\prime})t} \notag\\
&&\times \big[ \hat{a}_{\alpha\sigma}^{\dag}(E)\hat{a}_{\alpha\sigma}(E^{\prime})
-\hat{b}_{\alpha\sigma}^{\dag}(E)\hat{b}_{\alpha\sigma}(E^{\prime}) \big], \label{Jhatt}
\end{eqnarray}
which forms the basis for deriving two experimental observables: the period-averaged current $J$ and zero-frequency current noise $S$. 

Using Eqs.~\eqref{aa'} and \eqref{ba}, $\hat{J}(t)$ can be expressed as
\begin{eqnarray}
\hat{J}(t) =&& \frac{e}{2h}\sum_{\alpha}\sgn(\alpha) \int dE \int dE^{\prime} \sum_{\gamma\varsigma n}\sum_{\beta\chi m} e^{i(E-E^{\prime})t}p_{n,\gamma}^{\ast}  \notag\\
&&\hspace{-1.15cm}\times  p_{m,\beta}\hat{a}_{\gamma\varsigma}^{\prime\dag}(E-n\Omega) \hat{a}_{\beta\chi}^{\prime}(E^{\prime}-m\Omega) \big[ \tb{A}_{\gamma\beta}(\alpha,E,E^{\prime}) \big]_{\varsigma\chi}, \label{current}
\end{eqnarray}
where 
\begin{equation}
\tb{A}_{\gamma\beta}(\alpha,E,E^{\prime}) = \delta_{\alpha\gamma}\delta_{\alpha\beta} \tb{I} - \tb{s}^{\alpha\gamma\dag}(E) \tb{s}^{\alpha\beta}(E^{\prime}). \label{currentmatrix}
\end{equation}
For brevity, we denote $\tb{A}_{\gamma\beta}(\alpha,E,E)$ as $\tb{A}_{\gamma\beta}(\alpha,E)$ below.  

We now consider the period-averaged current:
\begin{equation}
J = \frac{1}{\mc T} \int_0^{\mc T} dt \, \langle \hat{J}(t) \rangle. \label{Javer}
\end{equation}
Using Eq.~\eqref{current}, the statistical average of $\hat{J}(t)$ reduces to the evaluation of the reservoir state averages:
\begin{eqnarray}
&&\quad \big\langle \hat{a}_{\gamma\varsigma}^{\prime\dag}(E-n\Omega) \hat{a}_{\beta\chi}^{\prime}(E'-m\Omega) \big\rangle \notag\\
&&= \delta_{\gamma\beta} \delta_{\varsigma\chi} \delta(E-n\Omega - E' + m\Omega) f_\gamma(E-n\Omega), \label{reser_average}
\end{eqnarray}
where $f_\gamma(E) = 1/\big[e^{(E-\sgn(\gamma)eV_\tr{dc})/k_B T}+1\big]$ is the Fermi distribution function of the $\gamma$-type carrier in the probe. Substituting Eqs.~\eqref{current} and \eqref{reser_average} into Eq.~\eqref{Javer} gives
\begin{equation}
J = \frac{e}{2h} \sum_{\alpha\gamma m} \sgn(\alpha) |p_{m,\gamma}|^2 \int dE \, \Tr[ \tb{A}_{\gamma\gamma}(\alpha,E) ] f_\gamma(E-m\Omega). \label{current2}
\end{equation}
Performing the explicit sum over $\alpha$ and $\gamma$ in Eq.~\eqref{current2} and using Eq.~\eqref{B4}, we obtain
\begin{eqnarray}
J = \frac{e}{h} \sum_m \int dE \, \Tr &&[ \tb{T}^{he}(E) ] \big[ |p_{m,e}|^2 f_e(E-m\Omega) \notag\\
&&- |p_{m,h}|^2 f_h(E-m\Omega) \big]. \label{current3}
\end{eqnarray}
Finally, making use of the relation $p_{m,h} = p^\ast_{-m,e}$, Eq.~\eqref{current3} immediately reduces to Eq.~\eqref{current4}.

\begin{widetext}
\section{Derivation of period-averaged zero-frequency current noise}\label{appc}
The time-symmetric current-current correlation operator is defined as
\begin{equation}
\hat{S}(t_{1},t_{2}) =  \Delta\hat{J}(t_{1})\Delta\hat{J}(t_{2}) 
+ \Delta\hat{J}(t_{2})\Delta\hat{J}(t_{1}) ,
\end{equation}
with $\Delta\hat{J}(t) = \hat{J}(t)-\langle \hat{J}(t)\rangle$. This symmetrization ensures that $\hat{S}(t_{1},t_{2})$ is Hermitian. The statistical average of $\hat{S}(t_{1},t_{2})$ is
\begin{equation}
S(t_{1},t_{2}) = \langle \{ \hat{J}(t_{1}), \hat{J}(t_{2}) \} \rangle 
- 2 \langle \hat{J}(t_{1}) \rangle \langle \hat{J}(t_{2}) \rangle, \label{St1t2new2}
\end{equation}
with $\{\cdot,\cdot\}$ denoting the anticommutator.
Inserting Eq.~\eqref{current} into Eq.~\eqref{St1t2new2} and making use of Wick’s theorem~\cite{haug2008quantum}, we find
\begin{eqnarray}
S(t_{1},t_{2})=&&\frac{e^{2}}{4h^{2}}\sum_{\alpha_{1}\alpha_{2}}\sum_{\gamma_{1}\delta_{1}}\sum_{m_{1}m_{2}n_{1}n_{2}}\sgn(\alpha_{1})
\sgn(\alpha_{2})p_{n_{1},\gamma_{1}}^{\ast}p_{m_{1},\delta_{1}}p_{n_{2},\delta_{1}}^{\ast}p_{m_{2},\gamma_{1}}\int dE_{1}\int dE_{1}^{\prime} e^{i(E_{1}-E_{1}^{\prime})(t_{1}-t_{2})}\notag\\
&& \times e^{i(n_{1}+n_{2}-m_{1}-m_{2})\Omega t_{2}}\Tr[ \tb{A}_{\gamma_{1}\delta_{1}}(\alpha_{1},E_{1},E_{1}^{\prime}) \tb{A}_{\delta_{1}\gamma_{1}}(\alpha_{2},E_{1}^{\prime}+(n_{2}-m_{1})\Omega,E_{1}+(m_{2}-n_{1})\Omega) ]\notag\\
&&\times\big\{f_{\gamma_{1}}(E_{1}-n_{1}\Omega) [1-f_{\delta_{1}}(E_{1}^{\prime}-m_{1}\Omega)]+f_{\delta_{1}}(E_{1}^{\prime}-m_{1}\Omega) [1-f_{\gamma_{1}}(E_{1}-n_{1}\Omega)]\big\} , \label{St1t2new}
\end{eqnarray}
where we have employed the statistical average in Eq.~\eqref{reser_average}.  

The period-averaged zero-frequency noise power $S$ is defined as~\cite{pedersen1998scattering}
\begin{eqnarray}
S = \frac{1}{\mc T} \int_{0}^{\mc T} dt \int d\tau S(t+\tau,t). \label{Saverage}
\end{eqnarray}
Substituting Eq.~\eqref{St1t2new} into Eq.~\eqref{Saverage} and making use of the integral relationships
\begin{equation}
\frac{1}{T}\int_{0}^{T} dt \, e^{i(n_{1}+n_{2}-m_{1}-m_{2})\Omega t} = \delta(n_{1}+n_{2}-m_{1}-m_{2}), \quad
\int d\tau \, e^{i(E_{1}-E_{1}^{\prime})\tau} = 2\pi \delta(E_{1}-E_{1}^{\prime}),
\end{equation}
we obtain
\begin{eqnarray}
S =&& \frac{e^{2}}{4h} \sum_{\alpha\beta\gamma\delta}\sum_{m_{1}m_{2}n_{1}} \sgn(\alpha)\sgn(\beta) p_{n_{1},\gamma}^{\ast} p_{m_{1},\delta} p_{m_{1}+m_{2},\delta}^{\ast} p_{m_{2}+n_{1},\gamma} \int dE \Tr[\tb{A}_{\gamma\delta}(\alpha,E) \tb{A}_{\delta\gamma}(\beta,E+m_{2}\Omega)]\notag\\
&&\times\big\{ f_{\gamma}(E-n_{1}\Omega)[1-f_{\delta}(E-m_{1}\Omega)]+ f_{\delta}(E-m_{1}\Omega)[1-f_{\gamma}(E-n_{1}\Omega)] \big\}.
\end{eqnarray}
Summing over $\alpha$, $\beta$, and $\delta$ yields $S=S_1+S_2$ with
\begin{eqnarray}
S_{1} && =\frac{e^{2}}{4h}\sum_{\gamma}\sum_{m_{1}m_{2}n_{1}}p_{n_{1},\gamma}^{\ast}p_{m_{1},\gamma}p_{m_{1}+m_{2},\gamma}^{\ast}p_{m_{2}+n_{1},\gamma
}\int dE \big\{f_{\gamma}(  E-n_{1}\Omega) [ 1-f_{\gamma}(E-m_{1}\Omega) ]+f_{\gamma}(  E-m_{1}\Omega) \notag\\
&&\qquad\times  [  1-f_{\gamma}(E-n_{1}\Omega) ]\big\}   \Tr[  ( \tb{A}_{\gamma\gamma}(  \gamma,E)  -\tb{A}_{\gamma\gamma}(  \bar{\gamma},E)  )  (  \tb{A}_{\gamma\gamma}(  \gamma,E+m_{2}\Omega)  -\tb{A}_{\gamma\gamma}(\bar{\gamma},E+m_{2}\Omega)  )  ],\label{S1}\\
S_{2} &&  =\frac{e^{2}}{4h}\sum_{\gamma}\sum_{m_{1}m_{2}n_{1}}p_{n_{1},\gamma}^{\ast}p_{m_{1},\bar{\gamma}}p_{m_{1}+m_{2},\bar{\gamma}}^{\ast}
p_{m_{2}+n_{1},\gamma}\int dE \big\{f_{\gamma}(  E-n_{1}\Omega) [ 1-f_{\bar{\gamma}}(E-m_{1}\Omega) ]+f_{\bar{\gamma}}(  E-m_{1}\Omega) \notag \\
&&\qquad\times [ 1-f_{\gamma}(  E-n_{1}\Omega)  ]\big\} \Tr[  ( \tb{A}_{\gamma\bar{\gamma}}(  \gamma,E)-\tb{A}_{\gamma\bar{\gamma}}(  \bar{\gamma},E)  )  (\tb{A}_{\bar{\gamma}\gamma}(  \gamma,E+m_{2}\Omega)
-\tb{A}_{\bar{\gamma}\gamma}(  \bar{\gamma},E+m_{2}\Omega))  ].\label{S2}
\end{eqnarray} 
The operators under the trace can be expressed explicitly making use of Eqs.~\eqref{B1}, \eqref{B2}, and \eqref{currentmatrix}:
\begin{eqnarray}
\tb{A}_{\gamma\gamma}(\gamma,E) - \tb{A}_{\gamma\gamma}(\bar{\gamma},E) &=& 2 \tb{T}^{\bar{\gamma}\gamma}(E), \label{S1trace} \\
\tb{A}_{\gamma\bar{\gamma}}(\gamma,E) - \tb{A}_{\gamma\bar{\gamma}}(\bar{\gamma},E) &=& 2 \tb{s}^{\bar{\gamma}\gamma\dag}(E) \tb{s}^{\bar{\gamma}\bar{\gamma}}(E), \label{S2trace1} \\
\tb{A}_{\bar{\gamma}\gamma}(\gamma,E) - \tb{A}_{\bar{\gamma}\gamma}(\bar{\gamma},E) &=& 2 \tb{s}^{\bar{\gamma}\bar{\gamma}\dag}(E) \tb{s}^{\bar{\gamma}\gamma}(E). \label{S2trace2}
\end{eqnarray}
Substituting Eq.~\eqref{S1trace} into Eq.~\eqref{S1} and summing over $\gamma$ leads to $S_1 = S_{1a} + S_{1b}$ with
\begin{eqnarray} 
S_{1a}=&&\frac{e^{2}}{h}\sum_{m_{1}m_{2}n_{1}}p_{n_{1},e}^{\ast}p_{m_{1},e}p_{m_{1}+m_{2},e}^{\ast}p_{m_{2}+n_{1},e}\int dE \Tr[  \tb{T}^{he}(  E)  \tb{T}^{he}(  E+m_{2}\Omega)  ]\notag\\
&&\times \big\{f_{e}(  E-n_{1}\Omega)  [ 1-f_{e}(  E-m_{1}\Omega) ] +f_{e}(  E-m_{1}\Omega) [1-f_{e}(  E-n_{1}\Omega) ]\big\}  ,\\
S_{1b}=&&\frac{e^{2}}{h}\sum_{m_{1}m_{2}n_{1}}p_{n_{1},h}^{\ast}p_{m_{1},h}p_{m_{1}+m_{2},h}^{\ast}p_{m_{2}+n_{1},h}\int dE \Tr[  \tb{T}^{eh}(  E)  \tb{T}^{eh}(  E+m_{2}\Omega)  ]\notag\\
&&\times \big\{f_{h}(  E-n_{1}\Omega)  [ 1-f_{h}(  E-m_{1}\Omega) ] +f_{h}(  E-m_{1}\Omega) [ 1-f_{h}(  E-n_{1}\Omega)]\big\}.
\end{eqnarray} 
Using the relations $p_{m,\alpha}^{\ast} = p_{-m,\bar{\alpha}}$, $1-f_{\alpha}(E) = f_{\bar{\alpha}}(-E)$, and Eq.~\eqref{B6}, one can readily find $S_{1b} = S_{1a}^{\ast}$, hence $S_1 = 2\Re S_{1a}$, yielding Eq.~\eqref{S1main}. Similarly, Eq.~\eqref{S2} reduces to Eq.~\eqref{S2main} making use of Eqs.~\eqref{S2trace1} and \eqref{S2trace2}.

\section{Excess photon-assisted noise of a normal metal-superconductor junction}\label{appd}
Our expressions Eqs.~\eqref{current4}--\eqref{S2main} are applicable to a normal metal-superconductor tunnel junction in the subgap transport regime where the bias amplitude is well below the superconducting gap. In the absence of subgap bound states, the scattering matrices are energy-independent, such that $\Tr[\tb{T}^{he}\tb{T}^{he}]=2\tau_A^2$ and $\Tr[\tb{s}^{he\dag} \tb{s}^{hh} \tb{s}^{hh\dag} \tb{s}^{he}]=2\tau_A(1-\tau_A)$, where $\tau_A$ denotes the Andreev reflection probability. We then obtain
\begin{eqnarray}
&&\hspace{-0.4cm}J=\frac{4e}{h}\tau_A\sum_n \vert p_{n,e}\vert^2 (eV_\tr{dc}+n\Omega)=\frac{4e^2}{h}\tau_A V_\tr{dc}, \label{J_finiteT}\\
&&\hspace{-0.4cm}S_{1}=\frac{8e^2}{h}\tau_{A}^{2}k_{B}T, \label{S1_finiteT}\\
&&\hspace{-0.4cm}S_{2}=\frac{4e^2}{h}\tau_{A}(1-\tau_{A})\sum_{n}P_n (2eV_\tr{dc}+n\Omega)\coth\frac{2eV_\tr{dc}+n\Omega}{2k_{B}T}, \label{S2_finiteT}
\end{eqnarray} 
where $P_n=\Big\vert\sum_{m}p_{n-m,e}(v)p_{m,e}(v)\Big\vert^{2}=\vert p_{n,e}(2v)\vert^2$. The resulting current $J$ and current noise $S\equiv S_1+S_2$ coincide with the expressions obtained via the Keldysh Green's function formalism~\cite{bertin2022microscopic}. 

In the zero-temperature limit, from Eqs.~\eqref{S1_finiteT} and \eqref{S2_finiteT} we obtain the excess noise, defined in Eq.~\eqref{excessnoise}, as
\begin{equation}
S^\tr{exc}=\frac{4e^2}{h}\tau_{A}(1-\tau_{A})\bigg[\sum_{n}P_n \vert2eV_\tr{dc}+n\Omega\vert-\vert2eV_\tr{dc}\vert\bigg], \label{dS_zeroT}
\end{equation}
which was previously derived within the framework of full counting statistics~\cite{belzig2016elementary}. When $eV_\tr{dc}/\Omega$ equals a half-integer, the excess noise exhibits a local minimum for periodic harmonic or square pulses, while it is fully suppressed (i.e., $S^\tr{exc}=0$) for periodic Lorentzian pulses~\cite{bertin2022microscopic}, reflecting the effective charge $2e$ of Cooper pairs. 

\section{Derivation of Eq.~\eqref{weakdS}}\label{appe}
When $v=eV_\tr{dc}/\Omega$ equals an integer $q$, variable substitutions $m \to m-q$ and $n \to n-q$ transform Eqs.~\eqref{weakS1} and \eqref{weakS2} into
\begin{eqnarray}
&&S_1 = \frac{2e^{2}}{h}\sum_{mn}X_{m-q,n-q}\bigg\vert \sum_{s=\pm}\int_{m\Omega}^{n\Omega}dE\,\tau_{s}^{2}(E)\bigg\vert, \label{appS1}\\
&&S_2 = \frac{2e^{2}}{h}\sum_{mn}X_{m-q,n-q}\bigg\vert \sum_{s=\pm}\int_{-m\Omega}^{n\Omega}dE\,\tau_{s}(E)\,[1-\tau_{s}(E)]\bigg\vert.\label{appS2}
\end{eqnarray}
For $\varepsilon_B=0$ and $\Gamma/\Omega\ll 1$, the integrands in Eqs.~\eqref{appS1} and \eqref{appS2} decay rapidly with $|E|$, which allows the integration bounds to be safely extended to $\pm\infty$ or $0$. The integral in Eq.~\eqref{appS1} is nonzero only when $mn\le0$, while that in Eq.~\eqref{appS2} is nonzero only when $mn\ge0$. Therefore,
\begin{eqnarray}
&&S_{1} = \frac{2e^{2}}{h}\Bigg[ Y_{1}\sum_{mn<0}X_{m-q,n-q}
+ \frac{Y_{1}}{2}\sum_{m\neq0,n=0}X_{m-q,n-q}
+ \frac{Y_{1}}{2}\sum_{m=0,n\neq0}X_{m-q,n-q}\Bigg],\\
&&S_{2} = \frac{2e^{2}}{h}\Bigg[ Y_{2}\sum_{mn>0}X_{m-q,n-q}
+ \frac{Y_{2}}{2}\sum_{m\neq0,n=0}X_{m-q,n-q}
+ \frac{Y_{2}}{2}\sum_{m=0,n\neq0}X_{m-q,n-q}\Bigg],
\end{eqnarray}
where $Y_1$ and $Y_2$ are defined in Eqs.~\eqref{Y1} and \eqref{Y2}.
When the ac bias component is switched off, $p_{m,\alpha}=\delta_{m0}$, leading to $S_1=0$ and $S_2=\frac{2e^{2}}{h}Y_{2}$ for $q \ne 0$. Consequently, noting that $\sum_{mn}X_{m-q,n-q}=1$, we obtain the excess noise 
\begin{equation}
S^\tr{exc}=\frac{2e^{2}}{h}\bigg\{  (  Y_{1}-Y_{2})  \bigg[\sum_{mn<0}X_{m-q,n-q}+\frac{1}{2}\sum_{mn=0}X_{m-q,n-q}\bigg]  -\frac{1}{2}(  Y_{1}+Y_{2})  X_{-q,-q}\bigg\}, \label{DeltaS1}
\end{equation}
which is presented as Eq.~\eqref{weakdS} in the main text.

\end{widetext}


%

\end{document}